# Bio-inspired Band Gap Engineering of Zinc Oxide by Intracrystalline Incorporation of Amino Acids

Anastasia Brif, Guy Ankonina, Christina Drathen and Boaz Pokroy*

Prof. B.Pokroy*, A. Brif

Department of Materials Science and Engineering and the Russell Berrie Nanotechnology Institute, Technion -Israel Institute of Technology, 32000 Haifa, Israel.

E-mail: bpokroy@technion.ac.il

Dr. G. Ankonina

Photovoltaic lab, Department of Electrical Engineering, Technion - Israel Institute of Technology, Haifa 32000, Israel.

Dr. C. Drathen

European Synchrotron Radiation Facility, BP 220, F-38043 Grenoble Cedex, France.

Crystal formation in biological systems has attracted many researchers over the years because of the enhanced structural properties—mechanical,[1] optical,[2] and magnetic[3]—of its outcome compared to non-biogenic crystals. Arguably by far the most research in the field has been focused on biogenic calcium carbonate and its properties.[4] Particular attention has been directed to biogenic calcite owing to its enhanced fracture toughness, shown to originate as a result of the deflection of propagating cracks away from the pronounced cleavage planes.[5] This is achieved by the presence of intracrystalline proteins and organic molecules within individual crystals in directions oblique to the (104) cleavage plane of calcite.[5a] These intracrystalline molecules have also been shown to strongly influence crystal shape and morphology[6] and coherence length.[7] Another outcome of their presence is the existence of systematic anisotropic lattice strains.[8] We have shown that incorporation of proteins and even of single amino acids into calcite that is grown synthetically leads to similar lattice strains.[9] Kim et al. showed that by mimicking proteins with micelles or polymer particles that become incorporated into calcite it is possible not only to reproduce lattice strains but also to enhance hardness of the calcite.[10] Similar results were obtained in a study by Schenk et al. using polyelectrolytes.[11] It was also shown that an agarose gel can be incorporated into single crystals of calcite.[12] Colfen et al. showed that amino acids affect the early stages of calcium carbonate formation.[13] A partial but deeper understanding of what governs the incorporation of biological molecules into calcium carbonate was recently achieved by the demonstration, after mapping of the incorporation of the 20 common amino acids into synthetic calcite.[9b]

Beyond the studies demonstrating that intracrystalline molecules exert certain effects on the structure of calcium carbonate, we know of no studies showing that amino acids or biological molecules can become incorporated into the lattice structures of other crystals. Here we report the unprecedented finding that, in a manner similar to what we observed in calcite, amino acids can be incorporated into the crystal lattice of ZnO. We show, moreover, not merely that such incorporation exists but also that the induced lattice strains accompanying the incorporation lead to systematic changes in the band gap of the semiconductor host.

Band gap engineering is of cardinal importance in a multitude of applications where a specific band gap value is needed,[15] for example in laser diodes, solar cells, heterojunction bipolar transistors, and many more. Tuning of the band gap can be achieved by one or more of the following methods:[15] (a) varying the chemical composition, (b) strain engineering via epitaxial crystal growth, and (c) size confinement on the nanometer scale. A linear relationship has been shown to exist between the band gap of semiconductors and lattice strain.[16] Here we not only studied the incorporation of the different amino acids into the lattice of ZnO by measuring the strain and performing chemical analyses, but we also investigated the effects of these intracrystalline amino acids on the band gap of the ZnO host. As shown below, those amino acids that were incorporated into the ZnO lattice also induced lattice strains that were accompanied by marked shifts in the band gap.

Our first objective was to determine whether amino acids would be able to become incorporated into the lattice of ZnO at all, as was previously shown for calcite.[9b] To this end we grew ZnO in the presence of the various amino acids at different solution concentrations. In the test case of serine incorporation, we found that serine indeed became incorporated and induced lattice strains of up to about 0.2% in the ZnO host. As the level of incorporation increased, so did the strain (see **Figure 1a**). Moreover, similarly to what we previously observed in biogenic calcite and biomimetic calcite, these strains relaxed upon mild thermal air annealing (300°C for 90 min) and a unique microstructure developed, characterized by broadening of the diffraction peaks (Figure 1b). Such broadening was unlike the pattern found under similar circumstances in most conventional materials (where the diffraction peaks are narrowed due to crystal growth and defect annealing), but was similar to what we previously observed in biogenic crystals.[17]

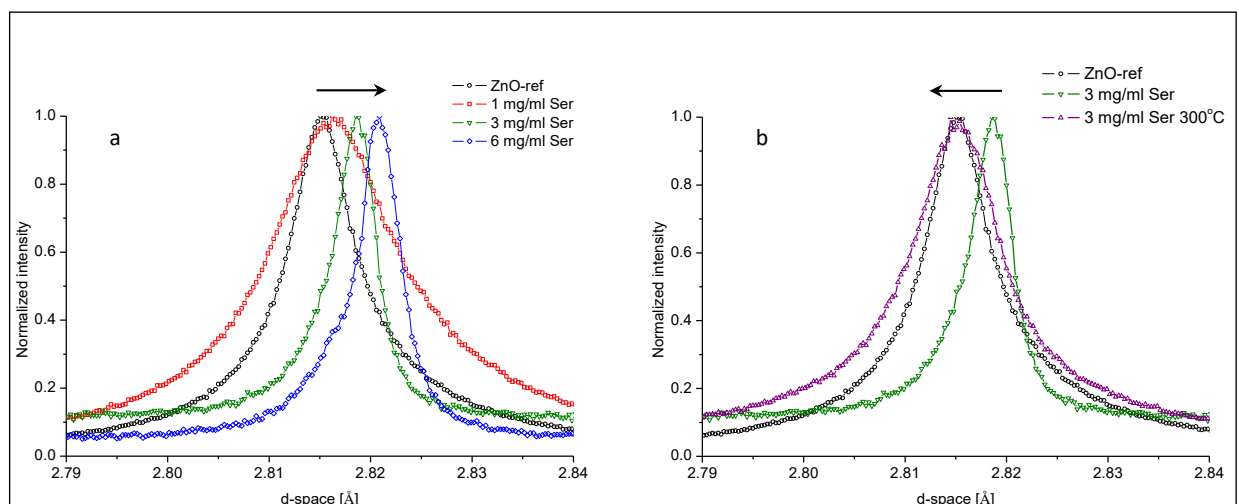

**Figure 1.** (100) XRD diffraction peak of ZnO crystals a) before and b) after thermal annealing of pure ZnO (dashed black) and ZnO grown in the presence of serine 1 mg/ml (red), 3 mg/ml (green), 3 mg/ml after annealing at 300ºC for 90 min (purple), and 6 mg/ml (blue).

We then screened all the common amino acids and found that a reasonable number of them indeed became incorporated at different levels into the ZnO lattice, with resulting lattice strains (**Figure 2a**). As with serine, all samples were also measured after mild heat treatment and all revealed lattice strain relaxation except cysteine and seleno-cysteine (see **Table S1**).

Those amino acids contain S and Se atoms which probably interact chemically with the ZnO matrix and therefore, higher annealing temperatures are required for lattice relaxation. Amino acid concentrations were estimated by X-ray photoemission spectroscopy (XPS) analysis to detect the atomic percentage of intracrystalline nitrogen (N) which corresponds to the intracrystalline amino acid concentration. XPS is a surface technique and provides information from only several nanometres below the surface but nevertheless we believe that it provides a very good estimate of the bulk amino acid incorporation levels. The latter assumption is based on the fact that for the case of Cysteine incorporation, we performed in addition energy dispersive spectroscopy (EDS) in the over a ZnO cross-section and found the S concentration to be very similar to that derived by XPS.

For each amino acid, the strain value was normalized by N at. % (see Figure 2b). Strain values for samples in which the nitrogen concentration was below the limit of detection (~0.1 at. %) were normalized to the detection limit concentration. No nitrogen was detected in the reference ZnO sample, which was grown in the absence of amino acids. From figure 2b we can conclude a strong correlation between shape change of the crystal and the measured strain, indicating that amino acids which strongly interact with the ZnO host, are more likely to be incorporated during the crystal growth. Such amino acids are those with electrically charged side chains (i.e. Lys, Asp, Glu), polar side chains (i.e. Ser, Tyr) and Cys and Sec. which probably form very strong bonds with Zn ions as we proposed previously for Ca in the case of calcite. [9b]

Interestingly, the highest incorporation level was shown in the case of serine and arginine. Using high-resolution scanning electron microscopy (HRSEM), significant changes in shape were observed when different amino acids were added. The resulting crystals could be divided into four different shape groups, described as flower, rounded flower, sphere, and rod-like shapes (Figure 2c to 2f). The resulting shape change was caused by different specific interactions of each amino acid with the ZnO. Similar shape changes have been described in the case of ZnO grown in the presence of vitamin C.[18] It has also been shown by Wenger et al. that latex particles (not single macromolecules) can get incorporated into the lattice of ZnO crystals and influence crystal morphology together with the optical and paramagnetic properties.[14]

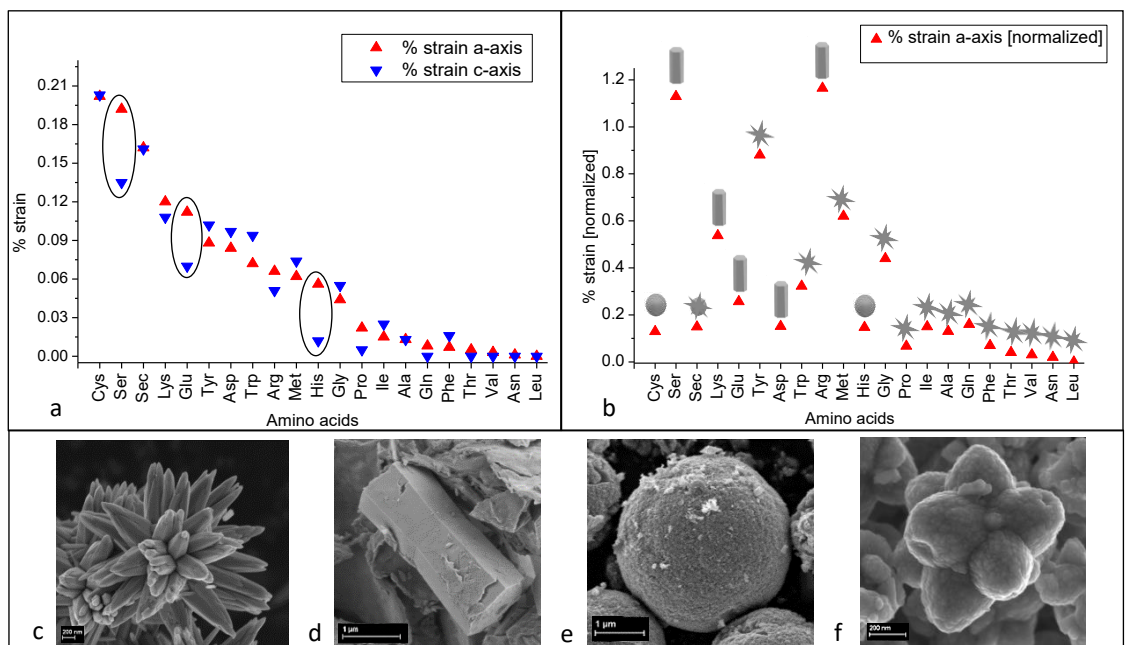

**Figure 2.** a) Lattice strain of ZnO with incorporated amino acids along the *a*-axis (red) and along the *c*-axis (blue). Circles indicate anisotropic strain. b) Lattice strain normalized by intracrystalline amino acid concentration. Schematics indicating crystal shape appear above each value. HRSEM images of crystal shapes for c) reference ZnO; and ZnO after incorporation with d) glutamic acid, e) cysteine, and f) selenocysteine.

These results clearly demonstrated that the incorporation of amino acids into other crystalline materials is feasible and might even be a widespread phenomenon. We believe that the findings of this study can have considerable impact on tuning the properties of new functional materials. In the present case, because ZnO is a semiconductor, we expected that the incorporation of organic molecules within the host lattice and the resulting lattice strain induction would alter the electronic properties as compared to pure ZnO crystals. We therefore measured the optical band gap of our amino acid-incorporating ZnO crystals. To quantify the band gap energy we obtained reflection spectra of the ZnO powders from diffused reflectance spectroscopy combined with the Kebulka-Munk (K-M) method.[19] According to the K-M method, the absorption coefficient α is defined as the ratio between the absorption coefficient $K$ and the scattering coefficient $S$ of a powder layer. This ratio can be calculated from the diffused reflectance $R_\infty$ by the equation: $\alpha = \dfrac{K}{S} = \dfrac{(1-R_\infty)^2}{2R_\infty}$ on the assumption that $R_\infty > 1$ if $K \neq 0$. Using the corresponding coefficient *n* associated with an electronic transition, a modified K-M function can be obtained from the expression $f(R) = (\alpha)^n$, while for direct band gap materials such as ZnO it is known that $n = 2$. To obtain the band gap energy for the powder samples we plotted the K-M modified function in energy units rather than in wavelength units. More details on extracting the band gap can be found in a recent work by López and Gómez.[20]

Using this method, we measured the changes in band gap energy (ΔEg) for different amino acid-incorporating ZnO crystals before and after thermal annealing. We did this for the various crystals possessing different levels of induced lattice strains. It was fascinating to observe, firstly, that the band gap was indeed altered due to the amino acid incorporation (**Figure 3a**), and secondly that there is a linear relationship between the magnitude of strain induced by the incorporation and the change in band gap of the ZnO crystalline host (Figure 3b).

To verify that the change in band gap was indeed caused by the incorporated amino acids, we measured the band gap on the same samples after mild thermal annealing, which also leads to full strain relaxation. We found that the band gap indeed returned to levels close to the control sample. In contrast, the band gap value of the reference ZnO remained almost the same after thermal annealing at 300°C. For the reference ZnO samples the average band gap energy was 3.28 eV, which corresponds with previously reported values of 3.2–3.4 eV.[21] ZnO samples incorporating cysteine exhibited the highest band gap increase, resulting in a band gap value of 3.41 eV, which was 4% higher than the band gap of the reference sample. For most of the amino acids we observed a positive linear correlation between band gap change and intracrystalline strain ($R^2 = 0.98$) (Figure 3b).

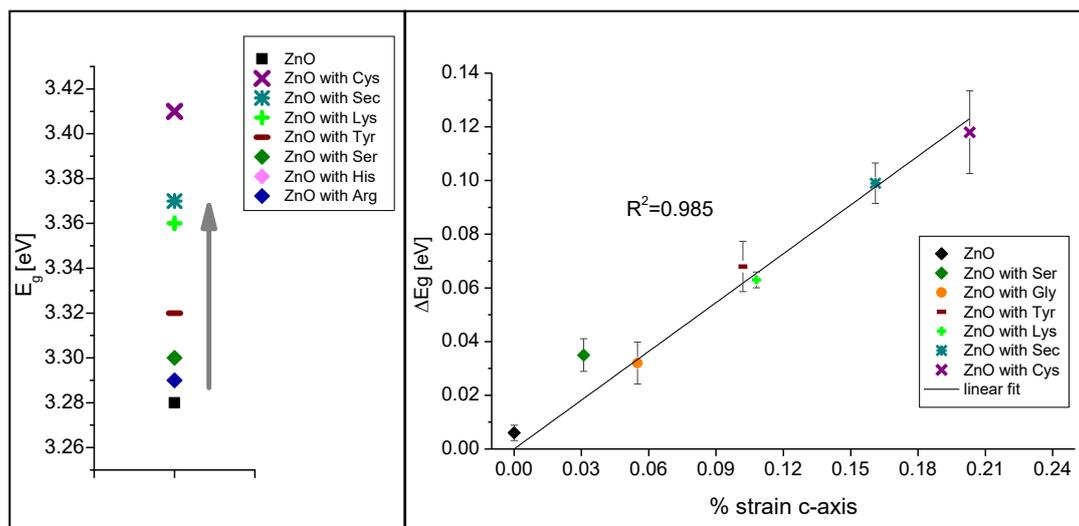

**Figure 3.** a) Band gap values and b) band gap energy change, $\Delta E_g$ (difference between before and after annealing) as a function of *c*-axis strain of ZnO samples crystallized in the presence of different amino acids.

The key experimental results obtained were (1) that amino acids can be incorporated into other inorganic crystalline hosts, in the present case specifically ZnO, and (2) that incorporation of amino acids into an inorganic crystalline semiconductor host induces not only lattice strains, as previously observed in biogenic and biomimetic calcium carbonate, but also a considerable band gap shift. In most of the tested amino acids the shift was an increase in band gap. This latter observation is most interesting, and to the best of our knowledge, unprecedented. We believe, moreover, that it could herald a new and bio-inspired route, in addition to and in combination with current methods, for tuning of the band gaps of semiconductors. In addition we call attention that there is a possibility that when semiconductors are grown in solution in the presence of organic molecules, some of these molecules unintentionally might get incorporated and alter the band gap.

With regard to the mechanism by which the band gap is altered, we should emphasize that we know of no other case in which amino acids interact with a semiconductor host to manipulate the band gap, and therefore that this finding will need more study in order to be clearly deciphered. Nevertheless, we can at this point clearly affirm that the state of lattice strain observed here is totally different from that seen in conventional epitaxial strain-induced band gap engineering. In the latter, a thin film is grown on a suitable substrate, which induces lattice strain via lattice mismatch. Owing to the Poisson effect the strain within the plane of the layer is opposite in sign to the strain perpendicular to it. This means that if the induced strain within the layer is positive, the perpendicular strain is negative. In ZnO, a positive strain along the *a*-axis leads to a decrease in the band gap.[22] The strain state in the present case is quite different, as along both the *a* and the *c* axes the strain is positive. In addition to this unexpected finding, it is important to remember that the strain is not mechanical like it is in the case of epitaxial strain. We cannot yet give a direct explanation for the observed band gap shift via the induced strain. To further examine the reason for this finding, it will be necessary to undertake an extensive study of the different molecular bond lengths in combination with atomistic modelling.

One possible explanation for the band gap shift might be related to the number of oxygen vacancies in the ZnO lattice. It has been shown that as the number of oxygen vacancies decreases the band gap increases.[23] Reaction of ZnO crystals with an oxidizing or a reducing environment proved, moreover, that this change was reversible.[24] It is feasible that the amino acids become incorporated within the ZnO lattice at oxygen-deficient locations, thereby effectively lowering the relative amount of oxygen vacancies, which in turn increases the band gap.

Another possible explanation for the shift in band gap is that the incorporated organics simply increase the effective dielectric constant of ZnO/amino acid composite crystals, which—at least theoretically—should also in turn increase the band gap.

We should re-emphasize, however, that at this point we cannot confirm the exact mechanism by which the incorporated amino acids alter the ZnO band gap, as more study will be needed in order to understand this phenomenon. We can conclude, however, that the results of this study clearly show that incorporated amino acids do lead to a rather strong shift (up to 4%) in the band gap of ZnO, and that in our opinion this bio-inspired route can serve as yet another avenue that can be added to or combined with the methods now commonly used for band gap engineering.

**Experimental Section**

*Materials:* ZnO crystals were crystallized from zinc nitrate hexahydrate (Scharlau Chemie, Spain) and ammonium hydroxide solution (Bio-Lab, Israel). Hydrochloric acid 37% (Merck) was used for pH control. The 21 amino acids used were L-aspartic acid (Asp), DL-tyrosine (Tyr), L-leucine (Leu), L-tryptophan (Trp), L-arginine (Arg), L-valine (Val), L-glutamic acid (Glu), L-methionine (Met), D-phenylalanine (Phe), DL-serine (Ser), D-alanine (Ala), L-glutamine (Gln), glycine (Gly), DL-proline (Pro), L-threonine (Thr), L-asparagine (Asn), L-lysine (Lys), L-histidine (His), L-isoleucine (Ile), L-cysteine (Cys), and Seleno-L-cysteine (Sec). All were purchased from Sigma-Aldrich. Deionized water (DI) was used for all the solutions.

*Crystal growth:* ZnO powders were precipitated from aqueous solution containing 0.25 M $Zn(NO_3)_2$ and pure ammonium hydroxide solution, in the presence of each of the 21 amino acids. Each amino acid was added with stirring to an aqueous solution of $Zn(NO_3)_2$ in the concentration range of 0.3 to 6 mg/ml. In all samples the pH was adjusted to 6 prior to reaction by addition of $NH_4OH$ or HCl. To initiate crystallization, 1 ml of $NH_4OH$ was added in drops to 100 ml of solution. Stirring was avoided to prevent early crystallization. The solution was transferred to a flask, immersed in a silicon oil bath, and kept at 95°C while stirring for 1 hour. The resulting ZnO powders were washed several times with DI water and air dried. Reference ZnO samples were prepared by the same method without amino acids.

*Characterization:* The powders were characterized by high-resolution powder X-ray diffraction (HR-XRD) utilizing a synchrotron source. Diffraction measurements were conducted on ID31 of the European Synchrotron Research Facility (ESRF), Grenoble, France, at a wavelength of 0.476798Å ± 0.000008Å. ZnO lattice parameter values were deduced by the Rietveld refinement method (using the GSAS program, EXPGUI interface). To determine the intracrystalline amino acid concentration, samples were analysed by XPS. Optical band gap was established by diffuse reflectance measurements using the Cary 5000 UV–Vis-NIR spectrophotometer (Agilent Technologies) with a DRA-2500 integrating sphere attachment. Optical reflection was obtained over a range of 250 to 800 nm.

**Supporting Information**
Supporting Information is available online from the Wiley Online Library or from the author.

**Acknowledgements**

The research leading to these results has received funding from the European Research Council under the European Union's Seventh Framework Program (FP/2013-2018) / ERC Grant Agreement n. [336077].

We thank Maria Koifman and Leonid Bloch for help in collecting and analyzing the XRD data at ID31 of the ESRF and Noga Kornblum for aid in part of the crystal synthesis. We are also indebted to the ESRF (ID31), and specifically to Dr. Andy Fitch, for use and support of the high-resolution powder beamline. We thank Prof. Emil Zolotoybko and Prof. Avner Rothschild for helpful discussions.

**Keyword** (bio-inspired crystal growth, lattice strain, ZnO, band gap engineering, biomineralization)

Anastasia Brif, Guy Ankonina, Christina Drathen and Boaz Pokroy*

**Title** Bio-inspired Band Gap Engineering of Zinc Oxide by Intracrystalline Incorporation of Amino Acids

ToC figure ((Please choose one size: 55 mm broad × 50 mm high **or** 110 mm broad × 20 mm high. Please do not use any other dimensions))

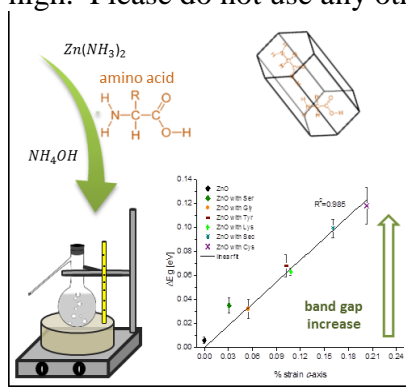

**Band gap engineering of Zinc Oxide semiconductors can be achieved using bio-inspired method.** During a bio-Inspired crystallization process, incorporation of amino acids into the crystal structure of ZnO induced lattice strain which induce linear band gap shifts. This allows for fine tuning of the band gap in a bio-inspired route.

# Supporting Information

for *Adv. Mater.*, DOI: 10.1002/adma.(201303596)

**Title**: Bio-inspired Band Gap Engineering of Zinc Oxide by Intracrystalline Incorporation of Amino Acids

Anastasia Brif, Guy Ankonina, Christina Drathen and Boaz Pokroy*

∗ **Table S1.** Quantitative data of amino acid content in the reaction solution, strain induced for each amino acid in both *a* and *c*- axes before and after thermal annealing (bold red), and the atomic percent (at. %) of intracrystalline amino acids found by dividing the N at. % by the number of N atoms in the molecule.

| Incorporated amino acid | Added amino acid concentration [mg/ml] | % strain *a*-axis | % strain *c*-axis | Rietveld refinement goodness of fit: $\chi^2$ | at. % of incorporated amino acid |
|---|---|---|---|---|---|
| **Asp** | 0.3 | 0.084<br>**0.001** | 0.097<br>**0** | 1.635<br>1.308 | 0.55 |
| **Glu** | 3 | 0.112<br>**0.019** | 0.07<br>**0.013** | 2.889<br>1.439 | 0.44 |
| **Cys** | 0.5 | 0.202<br>**0.243** | 0.203<br>**0.233** | 1.764<br>1.635 | 1.56 |
| **Sec** | 0.5 | 0.166<br>**0.099** | 0.161<br>**0.208** | 1.631<br>1.573 | 1.09 |
| **Ser** | 3 | 0.104<br>**0.007** | 0.031<br>**0.001** | 1.876<br>1.006 | 0.17 |
| **Lys** | 4 | 0.120<br>**0.009** | 0.108<br>**0.014** | 3.457<br>4.437 | 0.23 |
| **Gly** | 1 | 0.044<br>**0.007** | 0.055<br>**0.008** | 1.745<br>1.606 | <0.1 |
| **Tyr** | 0.5 | 0.088<br>**0.012** | 0.102<br>**0.008** | 2.304<br>1.324 | <0.1 |
| **Trp** | 1 | 0.072<br>**0.004** | 0.094<br>**0.004** | 1.935<br>1.26 | 0.23 |
| **Arg** | 4 | 0.066 | 0.051 | 1.42 | 0.06 |
| **His** | 1 | 0.056 | 0.012 | 4.667 | 0.38 |
| **Met** | 3 | 0.061 | 0.074 | 1.65 | <0.1 |
| **Pro** | 3 | 0.022 | 0.005 | 3.731 | 0.33 |
| **Ala** | 3 | 0.013 | 0.013 | 3.36 | <0.1 |
| **Leu** | 1 | 0.001 | 0.014 | 3.482 | <0.1 |
| **Ile** | 3 | 0.015 | 0.025 | 4.035 | <0.1 |
| **Val** | 3 | 0.003 | 0.042 | 4.574 | <0.1 |
| **Phe** | 3 | 0.007 | 0.016 | 3.824 | <0.1 |
| **Thr** | 1 | 0.005 | 0.011 | 4.091 | 0.125 |
| **Asn** | 1 | 0.001 | 0.002 | 3.043 | <0.1 |
| **Gln** | 1 | 0.008 | 0.05 | 4.878 | <0.1 |